\documentclass[%
 reprint,
 amsmath,amssymb,
 aps,
 prl, 
 floatfix,
]{revtex4-2}
\usepackage{svg}
\usepackage{graphicx}
\usepackage{dcolumn}
\usepackage{bm}

\begin{document}

\preprint{APS/123-QED}

\title{Dispersive wave propagation in disordered flexible fibers enhances stress attenuation}

\author{Peng Wang$^{1}$, Thomas Pähtz$^{2}$, Kun Luo$^{3}$, and Yu Guo$^{1,3,4,*}$}
\affiliation{
$^1$Department of Engineering Mechanics, Zhejiang University, Hangzhou, 310027, China\\
$^2$Institute of Port, Coastal and Offshore Engineering, Ocean College, Zhejiang University, Zhoushan, 316021, China\\
$^3$State Key Laboratory of Clean Energy Utilization, Zhejiang University, Hangzhou, 310027, China\\
$^4$Huanjiang Laboratory, Zhuji, Zhejiang Province, 311800, China\\
$^*$Corresponding author. Email: yguo@zju.edu.cn
}
\date{\today}
             
\begin{abstract}
We experimentally and computationally analyze impact-shock-induced stress wave propagation in packings of disordered flexible fibers. We find that dispersive wave propagation, associated with large stress attenuation, occurs much more prevalently in systems with larger fiber aspect ratios and moderate fiber flexibility. We trace these features to the microstructural properties of fiber contact chains and the energy-trapping abilities of deformable fibers. These findings provide new insights into the physics of the shock-impacted flexible fiber packings and open the way towards an improved granular-material-based damping technology. 
\end{abstract}
\maketitle

When a granular medium is hit by a shock, stress waves are generated and propagate within it through the network of interparticle contacts. Due to significant energy dissipation via the inelastic and sliding frictional contacts in this dynamic process, granular materials have been effectively used as dampers and protectors \cite{hidalgo2002,liu2019,chaunsali2017,khatri2011,zhang2021,kim2018,fraternali2009,sen2008,daraio2006,doney2006}. For a better understanding of the underlying granular physics, extensive studies have been performed on the wave propagation in granular media. The simplest such systems are one-dimensional (1D) granular chains composed of spheres \cite{liu2019}, cylinders \cite{chaunsali2017,khatri2011}, and more complex particles \cite{zhang2021,kim2018,fraternali2009}. Solitary waves are usually observed in the chains of monodisperse spherical particles \cite{sen2008}. If the spheres of different diameters or elastic moduli are introduced to the chains, wave reflection and decomposition can occur \cite{daraio2006,doney2006}. In the chains of identical cylindrical rods, nonlinear features of contacts and rod vibration lead to more localized solitary waves and stronger energy dissipation compared to the chains of spheres \cite{kim2015,xu2015}.

\begin{figure}[htbp]
\centering
\includegraphics[width=0.9\linewidth]{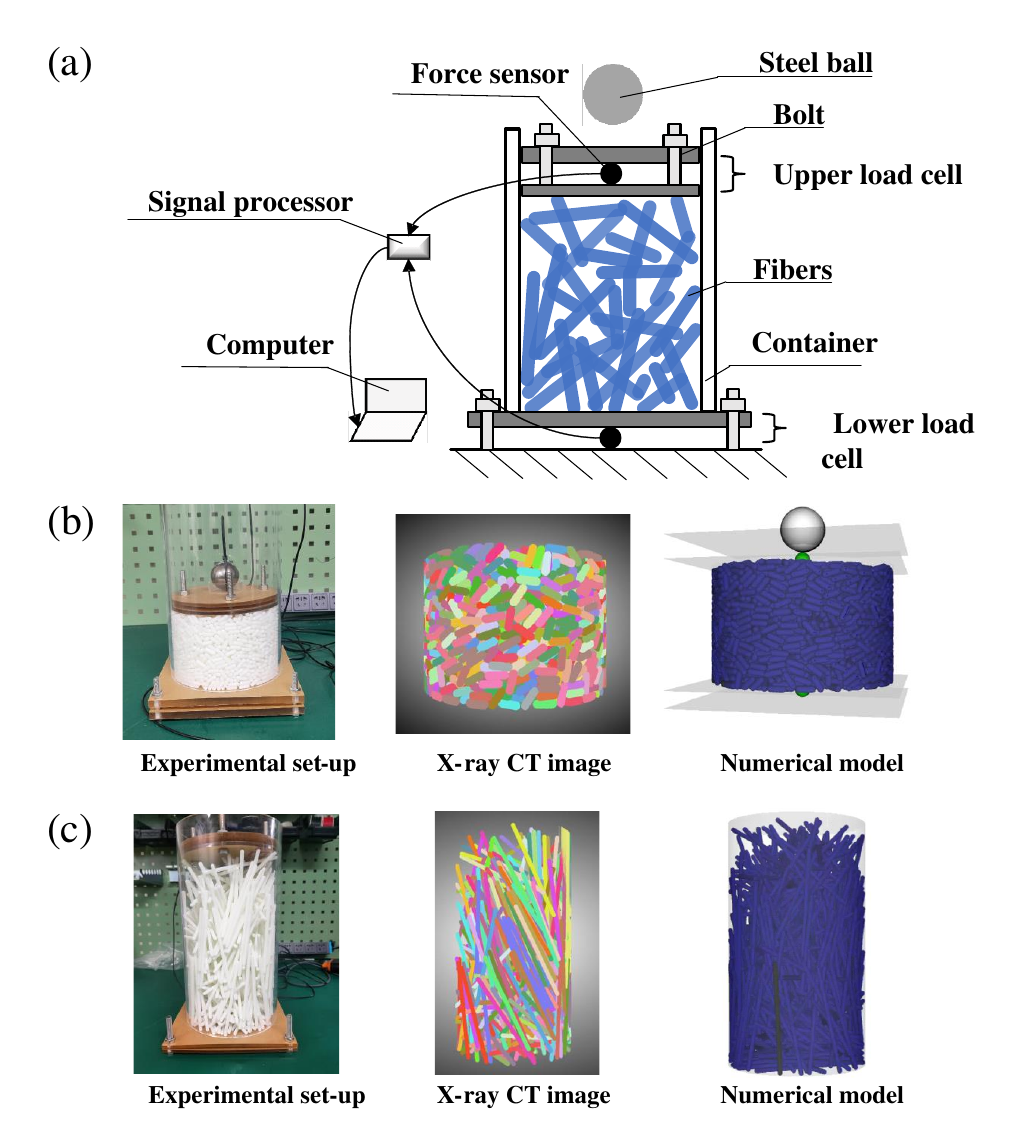}
\caption{\label{FIG1} (a) Schematic plot of experimental set-up of a free-falling steel ball impacting a bed of disordered fibers. Snapshot, X-ray CT image, and numerical model of the packed fibers with aspect ratios of (b) $AR$ = 3 and (c) $AR$ = 24.}
\end{figure}

In two-dimensional (2D) and three-dimensional (3D) granular systems, richer phenomena of wave propagation are observed and they show a strong dependence on the spatial arrangements and properties of particles \cite{wildenberg2013,kim2015solids,wang2019,wang2022}. Even a slight variation in packing structure, such as the presence of small gaps due to particle size tolerance in 2D ordered granular media, can significantly influence wave speeds and peak forces, causing scatter in experimental results \cite{waymel2018}. Varying number ratio or stiffness ratio of two granular components can produce diverse wave speeds \cite{taghizadeh2023} and wave fronts \cite{leonard2012} by redirecting energy transmission within the system, demonstrating a wave-control approach through manipulating composition and topological structure of the granular system.

Compared to systems of rigid spheres and short cylinders, stress wave propagation is much less understood in 3D assemblies of disordered flexible fibers, which have practical applications in lightweight building materials, proppant in oil recovery, and colloidal and granular polymers \cite{desmond2006,hoy2017}. The fiber systems have complex microstructures with interlocking particle contacts and large particle deformation \cite{desmond2006,hoy2017,guerra2023}, which substantially complicate energy transmission and thus alter the ability of stress attenuation. 

In this Letter, we find, in experiments and numerical simulations of shock impacts on assemblies of disordered flexible fibers, that the occurrence of solitary or dispersive wave propagation depends on fiber aspect ratio ($AR$) and flexibility. Stress wave attenuation is remarkably enhanced by dispersive pattern of wave propagation, in which energy trapping due to fiber deformation plays an important role. Thus, extent of the stress attenuation and wave speed can be modulated by adjusting fiber properties, providing a new wave-control and damping technology based on the flexible fibers. 

In our experiments, as shown in Fig. 1(a), fibers are randomly packed in a cylindrical container and sandwiched between two load cells, on which force sensors are installed. A steel ball of 4 cm in diameter falls under gravity and impacts on the upper load cell at a velocity of 0.4 m/s, generating a shock wave that propagates through the fiber bed from the top to the bottom. The time evolution of the forces on the upper and lower load cells is automatically recorded by a computer system. 

The fibers of various $AR$s are made of photopolymer resin and fabricated using a 3D printing machine. The morphology of the fiber beds resolved to single-fiber scale is obtained using a 320 kV X-ray Computed Tomography (CT) in-situ testing instrument (see Figs. 1(b) and 1(c)). Utilizing the X-ray CT data, high-fidelity numerical models of the fiber beds, which have the same morphology to the experimental set-ups, can be generated for numerical simulation studies using the Discrete Element Method (DEM). In the DEM simulations, a semi-flexible fiber is represented by a string of bonded sphero-cylinders, and stretching/compression, bending, and twisting deformations of the fiber are governed by a set of elastic constitutive laws \cite{guo2018,guo2020}. The interactions between fiber-fiber, fiber-plane and fiber-cylindrical boundary are modelled as normal and tangential contact forces, described by the modified Hertz-Mindlin models \cite{guo2020}. Viscous dissipation of energies in the contacts and fiber deformations is considered through contact damping and bond damping forces in the DEM simulations \cite{guo2020}. Additional information about the experimental setup, X-ray CT measurement, and DEM theories is provided in Supplementary Material \cite{supplemental}.

In the experiments, the forces exerted on the upper load cell, referred to as source, and the lower load cell, as receiver, are recorded as a function of time (Fig. 2(a)). The peak force on the receiver $F_R^p$ is remarkably smaller than that on the source $F_S^p$, indicating significant attenuation in force transmission through the fiber bed. Such force attenuation was also observed in a medium of rigid spheres attributed to Rayleigh scattering and contact damping \cite{zhai2020}. As $AR$ increases, the receiver force lasts for a longer time with a decrease in the peak $F_R^p$. In addition, a smaller force pulse on the source is induced by post-shock expansion of the fiber bed at a late time for $AR$ = 24. The similar force attenuation behaviors are obtained for various $AR$s in the DEM simulations (Fig. 2(b)).

\begin{figure*}[htbp]
\centering
\includegraphics[width=0.82\linewidth]{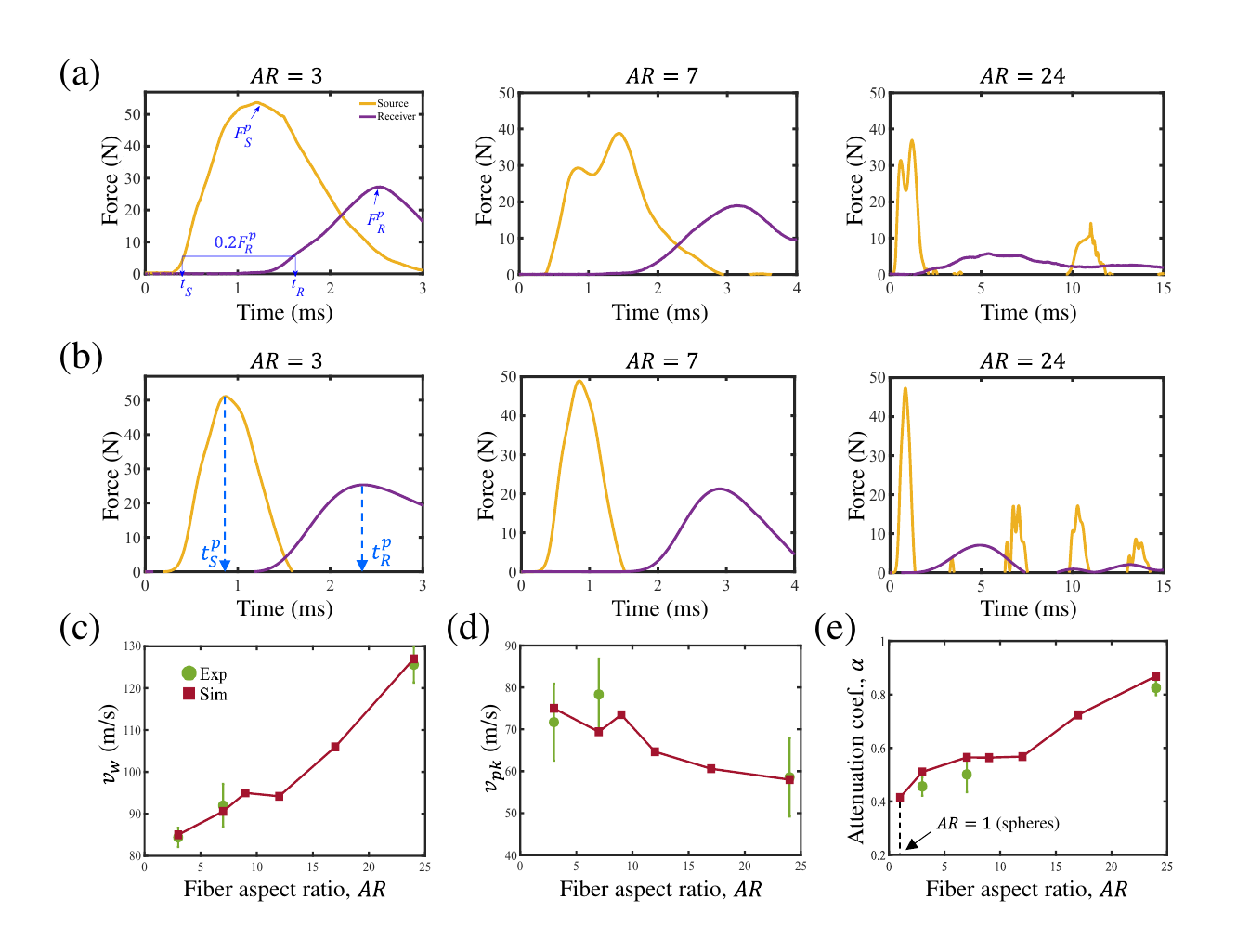}
\caption{\label{FIG2} (a) Experimental and (b) DEM simulation results of forces exerted on the upper load cell (referred to as source) and lower load cell (receiver) as a function of time for the fiber beds with fiber aspect ratios of $AR$=3, 7, and 24. Wave speed $v_w$, speed of peak force $v_{pk}$, and attenuation coefficient $\alpha$ are plotted as a function of $AR$ in (c), (d), and (e), respectively. Error bars represent SDs across ﬁve measurements.}
\end{figure*}

The time instants at which the source and receiver forces reach 20\% of the peak receiver force $F_R^p$ (i.e. ${0.2F}_R^p$) are written as $t_S$ and $t_R$, respectively, as illustrated in Fig.2(a). We define the wave speed $v_w$ as
\begin{equation}
    v_w = \frac{H}{t_R - t_S},
\end{equation}
in which $H$ is the distance between the upper and lower load cells. The time instants $t_S^p$ and $t_R^p$ (Fig.2(b)) are corresponding to the peak forces on the source and receiver, respectively. Thus, the speed of peak force $v_{pk}$ is defined as 

\begin{equation}
    v_{pk} = \frac{H}{t_R^{p} - t_S^{p}}.
\end{equation}

With increasing $AR$, the wave speed $v_w$ increases (Fig.2(c)), the speed of peak force $v_{pk}$ decreases (Fig.2(d)), and the attenuation coefficient, defined as $\alpha = (F_S^p - F_R^p) / F_S^p$, increases (Fig.2(e)). These trends are determined by microstructural properties of force transmission paths and mechanism of energy propagation and dissipation, which will be discussed later. 

To explore the process of stress wave propagation, the fiber bed is equally partitioned into 20 layers in the DEM simulations. Total fiber-fiber contact force $F_c$, total kinetic energy $K$ and total potential energy $U$ (due to elastic deformation) of the fibers in each layer can be calculated for a specified time instant. Treating the initial quantities before the impact ($F_c^0$, $K^0$, and $U^0$) as the benchmarks, the corresponding incremental contact force $\Delta F_c=F_c-F_c^0$, kinetic energy $\Delta K=K-K^0$, and potential energy $\Delta U=U-U^0$ reflect the changes made by the stress wave. 

Spatiotemporal distributions of $\Delta F_c$, $\Delta K$, and $\Delta U$ are plotted in Fig.3, and different patterns of wave propagation are observed for the fiber beds with different fiber aspect ratios. For the short fibers of $AR$ = 3, an inclined, narrow band of $\Delta F_c$ is formed at early time (Fig.3(a)), demonstrating solitary wave propagation. The contact forces drive the movement and deformation of the fibers. Thus, the incremental kinetic energy $\Delta K$ and potential energy $\Delta U$ transmit in a similar manner as $\Delta F_c$ (Figs.3(d) and (g)). Nevertheless, $\Delta K$ lasts longer in the upper region than in the lower region of the bed (Fig.3(d)). The wave propagation pattern with $AR$ =7 (Figs.3(b), (e), and (h)) resembles that with $AR$ = 3 (Figs.3(a), (d), and (g)), except that stronger attenuation with smaller $\Delta F_c$ and $\Delta K$ towards the bottom is obtained for $AR$ = 7 compared to $AR$ = 3. It is observed that the potential energy $\Delta U$ for $AR$ = 7 is two-order larger than that for $AR$ = 3, attributed to the larger bending deformation of the larger $AR$ fibers. Thus, as $AR$ increases, more energy is absorbed by the fiber deformation and converted to the potential energy, preventing the propagation of force and energy to the lower region.

Unlike the solitary waves in the short fiber beds with $AR$ = 3 and 7, in the beds of a much larger aspect ratio $AR$ = 24, stress wave propagates dispersedly with significant contact forces $\Delta F_c$ lasting for a much longer duration and remaining primarily in the upper region of the bed (Fig.3(c)). Consistently, the energies $\Delta K$ and $\Delta U$ are trapped in the upper region and hardly propagate to the bottom (Figs.3(f) and (i)). As a result, lower speeds of the peak force $v_{pk}$ and larger attenuation $\alpha$ are obtained for the larger fiber aspect ratios (Figs.2(d) and (e)), due to the dispersive wave propagation.

\begin{figure*}[htbp]
\centering
\includegraphics[width=0.82\linewidth]{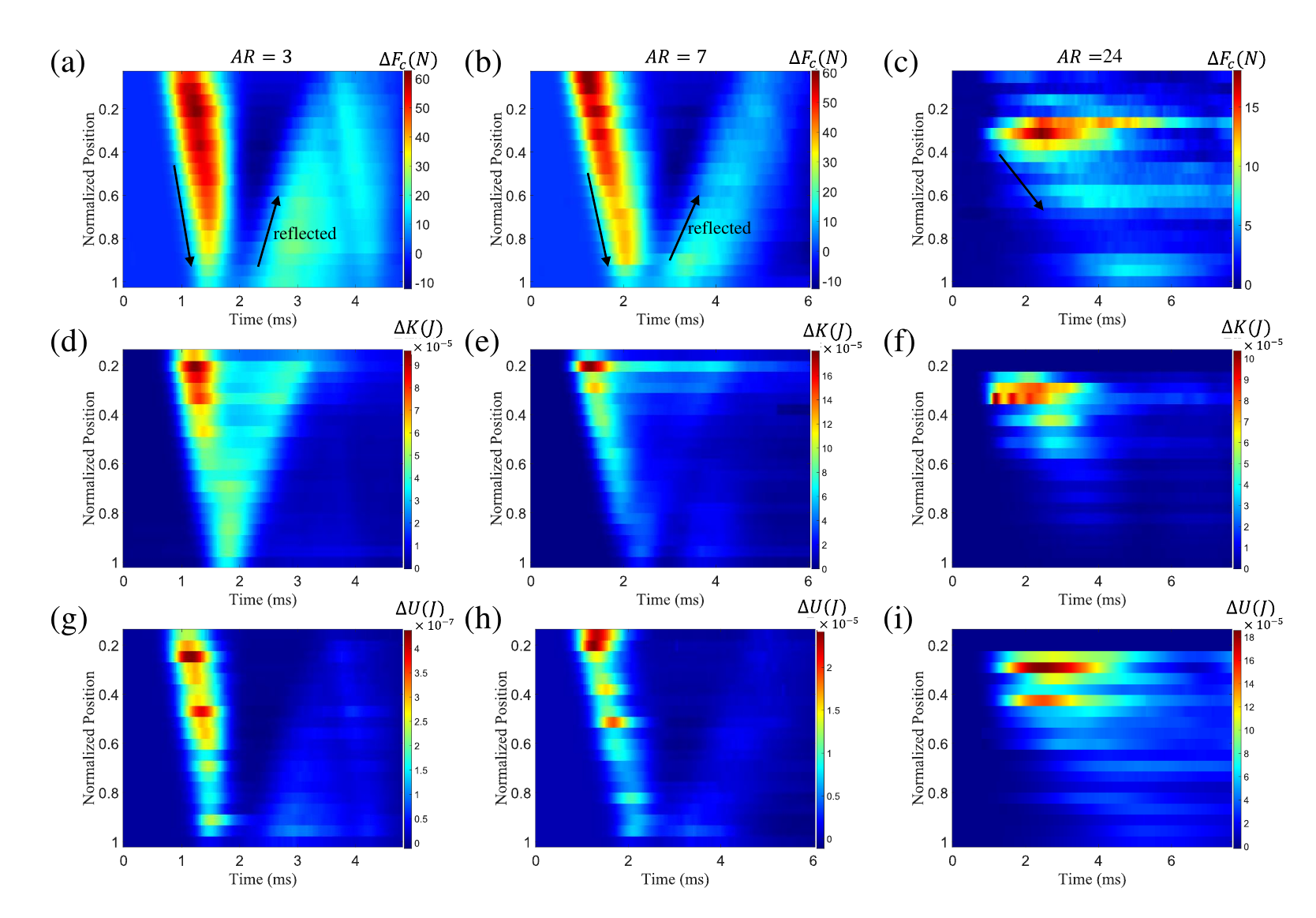}
\caption{\label{FIG3} Spatiotemporal distributions of (a-c) incremental contact force $\Delta F_c$, (d-f) incremental kinetic energy  $\Delta K$, and (g-i) incremental potential energy  $\Delta U$ for the fiber beds with fiber aspect ratios of $AR$=3, 7, and 24. The present results are obtained from the DEM simulations.}
\end{figure*}

Morphology of the shortest percolation force chain, which is the shortest path between the source and receiver plates in the network of force chains \cite{bhosale2022,shukla1991,howell1999,owens2011}, characterizes the microstructure and connectivity of a granular system \cite{taghizadeh2023} and therefore has a crucial impact on the stress wave propagation. Based on the X-ray CT data of the packed fiber beds in the experiments, the shortest percolation force chains are determined (via the algorithm provided in Supplementary Material \cite{supplemental}) for the beds with various $AR$s, as highlighted by the red particles that construct the chains in Fig.4(a). It is observed that the number of fibers in the shortest force chain $N$ decreases with increasing $AR$ (Fig.4(b)). Assuming the average tilting angle of the fibers from the horizontal plane as $\theta$, the projection of a single fiber in the direction of wave propagation (vertical in the present setups) is $AR\cdot d_f\cdot\sin{\theta}$, in which $d_f$ is fiber diameter. Thus, the smallest number of fibers in the shortest force chain can be written as
\begin{equation}
        N=\frac{H_0}{AR\cdot d_f\cdot sin\theta\ }
\end{equation}
in which $H_0$ is the fiber bed height between the source and receiver plates and has a correlation with packing density $\phi$,
\begin{equation}
   H_0=\frac{V_f^{total}}{A_S\cdot\phi\ } ,                          
\end{equation}
where $V_f^{total}$ is the total volume of the fibers in the bed and $A_S$ is the cross sectional area of the cylindrical container. Substituting Eq.(4) to (3), it gives 
\begin{equation}
    N=\frac{V_f^{total}}{A_s\cdot d_f\cdot AR\cdot\phi\cdot sin\theta} .                  
\end{equation}

For randomly-packed, monodisperse rods, the correlation between the packing density $\phi$ and $AR$ is provided in \cite{desmond2006}. In addition, the average tilting angle of the fibers $\theta $ is also a function of $AR$, as presented in Supplementary Material \cite{supplemental}. Therefore, according to Eq.(5), $N$ can be expressed as a function of $AR$ in Fig.4(b). In a chain of more fibers, the force pulse needs to transmit through more fiber-fiber contact points, slowing down transmission speed due to small contact areas and dissipative nature of the contacts. Thus, lower wave speeds $v_w$ are obtained for the beds with smaller $AR$s (Fig.2(c)), which have larger numbers of fibers in their shortest force chains (Fig.4(b)). 

\begin{figure}[htbp]
\centering
\includegraphics[width=0.8\linewidth]{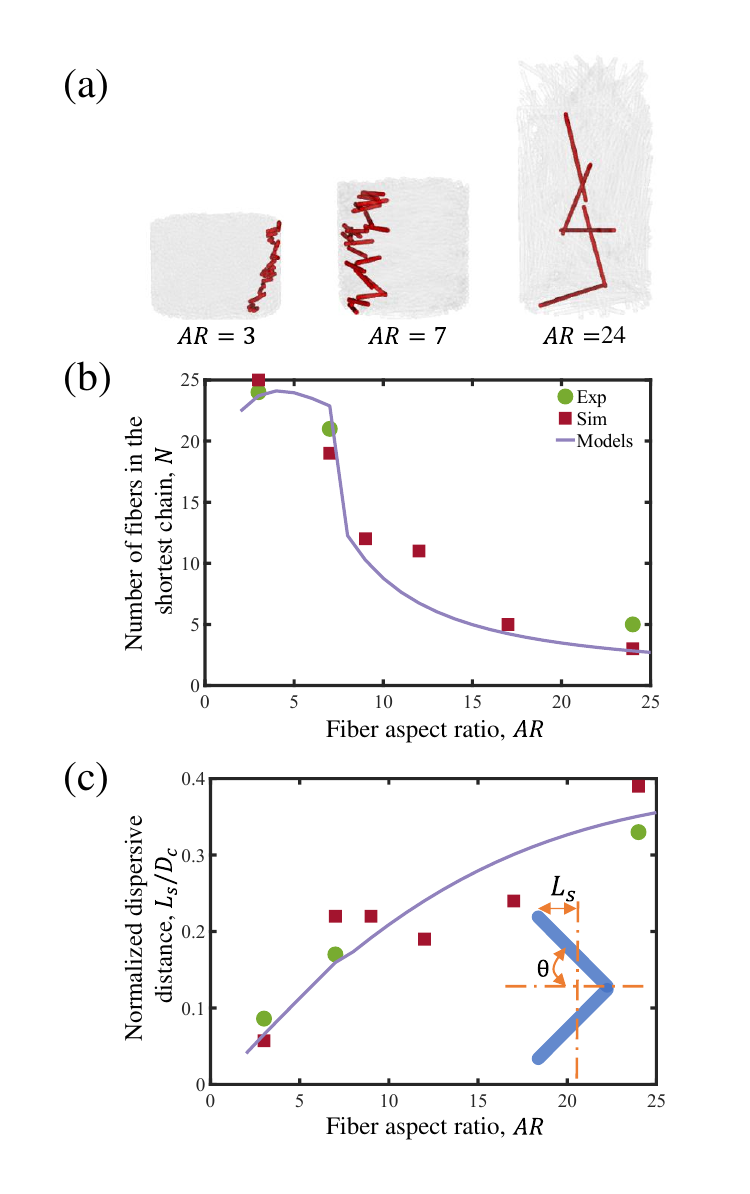}
\caption{\label{FIG4} (a) Shortest percolation force chains obtained from the X-ray CT images in the experiments. (b) Number of fibers in the shortest force chain $N$ and (c) normalized dispersive distance $L_s/D_c$ as a function of fiber aspect ratio $AR$.}
\end{figure}
Spatial deviation of a force chain from the propagation direction plays a significant role in dispersing the wave and dissipating the energy. As shown in Fig.4(c), a dispersive distance $L_s$ is defined as the half length of the projection of a fiber in the transverse (horizontal in the present setups) direction, 
\begin{equation}
     L_s=\frac{1}{2}AR\cdot d_f\cdot\cos{\theta},                    
\end{equation}
which is normalized by the diameter of the cylindrical container $D_c$, 
\begin{equation}
    \frac{L_s}{D_c}=\frac{AR\cdot d_f\cdot\cos{\theta}}{2D_c}.
\end{equation}
As $AR$ increases, the normalized dispersive distance $L_s/D_c$ increases (Fig.4(c)), enhancing dispersive propagation of force and energies (Figs.3(c),(f), and (i)).

\begin{figure}[htbp]
\centering
\includegraphics[width=\linewidth]{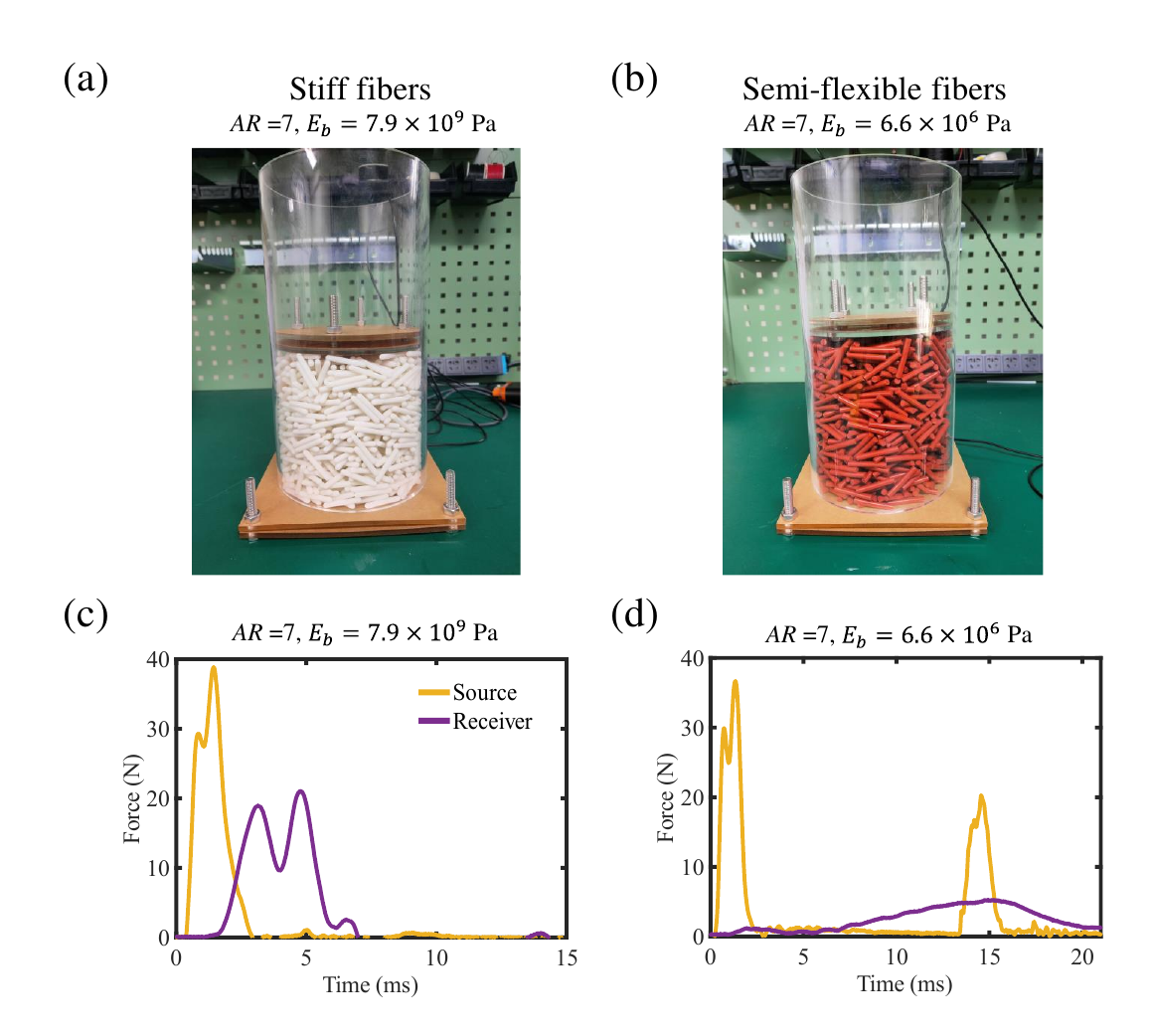}
\caption{\label{FIG5} Experimental setups of impacts on (a) stiff and (b) semi-flexible fibers. Experimental results of forces exerted on the source and receiver as a function of time for the beds of (c) stiff and (d) semi-flexible fibers.}
\end{figure}

Flexibility of fibers is characterized by fiber bending modulus $E_b$, quantifying the capacity to resist fiber bending deformation (see Supplementary Material \cite{supplemental}). The fibers are more flexible with a smaller $E_b$. In our experiments, stiff fibers with $E_b=7.9\times{10}^9$ Pa (made of photopolymer resin) and semi-flexible fibers with $E_b=6.6\times{10}^6$ Pa (silicone rubber) are used to investigate the effect of fiber flexibility on the wave propagation (Figs.5(a) and (b)). More significant attenuation with a much smaller force on the receiver is obtained for the semi-flexible fibers (Fig.5(d)) than the stiff ones (Fig.5(c)). It takes much longer time for the receiver force to reach the peak for the semi-flexible fibers, and meanwhile a second force impulse on the source plate occurs, due to the post-shock expansion of the fiber bed (Fig.5(d)). These results demonstrate the excellent ability of energy absorption and shock mitigation for the semi-flexible fibers. 

The DEM simulations allow us to check the effect of fiber bend moduli $E_b$ spanning a wide range of 10 to ${10}^{10}$ Pa. As shown in Fig.6, the largest attenuation coefficients $\alpha$ and lowest wave speeds $v_w$ are obtained in the beds of semi-flexible fibers with $10^2  \text{Pa} < E_b < 10^7  \text{Pa}$, through which wave propagates dispersedly as depicted in the spatiotemporal distribution of $\Delta F_C$ for $E_b=6.6\times{10}^6$ Pa. The dispersive wave propagation is attributed to the fact that more energy is trapped in the form of potential energy due to fiber deformation (Supplementary Material \cite{supplemental}). As the fibers become stiffer, i.e. $E_b>{10}^8$ Pa, the attenuation $\alpha$ decreases and wave speed $v_w$ increases. The small fiber deformation results in a low capacity to store the potential energy. In addition, the wave propagates faster within stiff fibers. Thus, solitary wave is observed in the spatiotemporal distribution of $\Delta F_c$ for the stiff fibers. 

\begin{figure*}[htbp]
\centering
\includegraphics[width=0.9\linewidth]{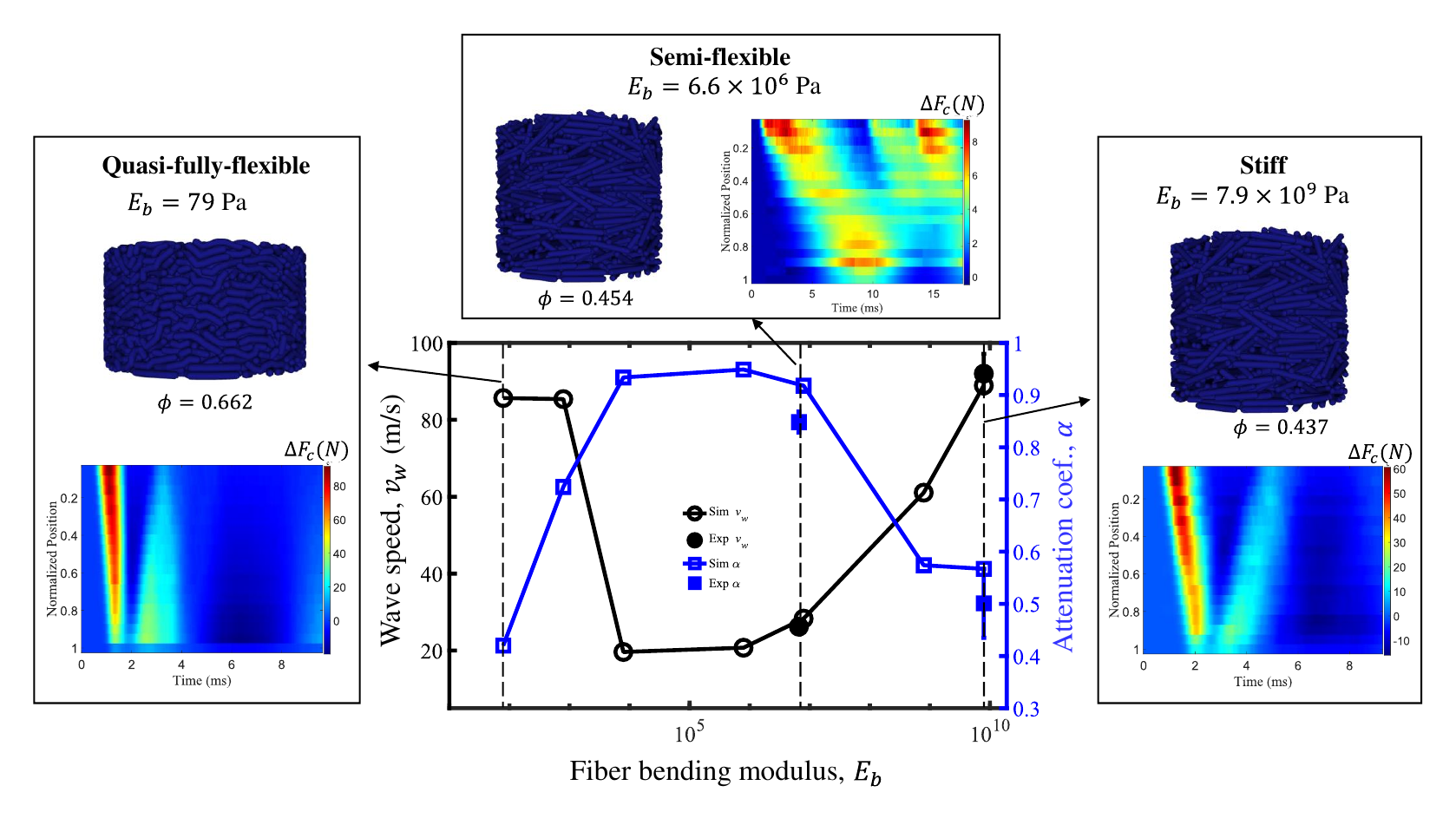}
\caption{\label{FIG6} Wave speed $v_w$ and attenuation coefficient $\alpha$ as a function of fiber bending modulus $E_b$. The inserts show the snapshots of fiber beds and spatiotemporal distributions of the incremental contact force $\Delta F_c$ obtained from the DEM simulations.}
\end{figure*}

When the fibers are nearly completely flexible with $E_b<{10}^3$Pa, the bed is consolidated with a large packing density ($\phi=0.662$ for $E_b$=79 Pa). Strong contact force chains in a densely packed bed facilitate the force and energy transmission, leading to higher wave speeds $v_w$ (Fig.6). In spite of large fiber deformation, the quasi-fully-flexible fibers have a weak capacity to store the potential energy (Supplementary Material \cite{supplemental}), due to the very small bending modulus $E_b$. Thus, solitary wave propagation is obtained in the quasi-fully-flexible fibers, similar to the wave pattern in the stiff fibers, and the force pulse is less reduced with smaller attenuation coefficients $\alpha$. As a result, it is found that the semi-flexible fibers with moderate bending moduli have the best ability to absorb energy and reduce the shock impact. 

In conclusion, we experimentally and computationally elucidate stress wave propagation in packings of disordered flexible fibers. We find that dispersive wave propagation contributes to a large extent of attenuation in forces and energies. A larger fiber aspect ratio leads to fewer fibers in the shortest percolation force chain, increasing wave speed, and a larger dispersive distance of the force chain, promoting the dispersive wave propagation. Semi-flexible fibers with moderate fiber bending moduli $E_b$ have the best capacity to absorb kinetic energy by converting it to potential energy through fiber deformation, which is eventually dissipated through inter-fiber contacts and fiber vibration. Consequently, lower wave speeds and larger attenuation in transmitted forces are obtained for semi-flexible fibers compared with stiff fibers and quasi-fully-flexible fibers. The findings in this study open new possibilities for wave-control technology and improved design of energy-absorbing and shock-proof materials.

 This work was financially supported by the National Natural Science Foundation of China (NOs. 12372250 and 12132015), Zhejiang Provincial Natural Science Foundation of China (NO. LZ24A020002), and Specialized Research Projects of Huanjiang Laboratory (NO. XYY-128102-E52201).

\nocite{*}

%

\end{document}